\documentclass{article} %
\usepackage{iclr2025_conference,times}

\usepackage{cite}
\usepackage{soul}
\usepackage{amsmath,amssymb,amsfonts}
\usepackage{algorithmic}
\usepackage{graphicx}
\usepackage{textcomp}
\usepackage{xcolor}
\usepackage{microtype}
\usepackage{listings}
\usepackage[hyphens]{url}
\usepackage{hyperref}
\usepackage{graphicx} 
\def\BibTeX{{\rm B\kern-.05em{\sc i\kern-.025em b}\kern-.08em
    T\kern-.1667em\lower.7ex\hbox{E}\kern-.125emX}}

\definecolor{lightyellow}{RGB}{252,244,133} \definecolor{darkyellow}{RGB}{255,209,0}
\definecolor{lightRed}{RGB}{242,220,218} \definecolor{darkRed}{RGB}{217,150,144}
\definecolor{lightPurple}{RGB}{221,215,230} \definecolor{darkPurple}{RGB}{205,194,217}
\definecolor{lightGreen}{RGB}{235,241,223} \definecolor{darkGreen}{RGB}{205,220,175}
\definecolor{lightBlue}{RGB}{219,238,244} \definecolor{darkBlue}{RGB}{147,205,221}

\DeclareRobustCommand{\rone}[1]{{\sethlcolor{darkGreen}\hl{#1}}}

\DeclareRobustCommand{\rthree}[1]{{\sethlcolor{darkPurple}\hl{#1}}}
\DeclareRobustCommand{\rfour}[1]{{\sethlcolor{darkRed}\hl{#1}}}

\DeclareRobustCommand{\rone}[1]{#1}

\DeclareRobustCommand{\rthree}[1]{#1}
\DeclareRobustCommand{\rfour}[1]{#1}

\author{Riselda Kodra \\  
Swiss Federal Institute of Technology \\
1008 Lausanne, Switzerland \\
\texttt{riselda.kodra@epfl.ch} \\
\AND
Hadjer Benmeziane, Irem Boybat, William Andrew Simon \\
IBM Research Zurich \\
8803 Rüschlikon, Switzerland \\
\texttt{\{hadjer.benmeziane, ibo, william.simon1\}@ibm.com}
}

\iclrfinalcopy %
\begin{document}

\title{Enhancing Downstream Analysis in Genome Sequencing: 
Species Classification While Basecalling}
 
\maketitle

\begin{abstract}
The ability to quickly and accurately identify microbial species in a sample, known as metagenomic profiling, is critical across various fields, from healthcare to environmental science. This paper introduces a novel method to profile signals coming from sequencing devices in parallel with determining their nucleotide sequences, a process known as basecalling, via a multi-objective deep neural network for simultaneous basecalling and multi-class genome classification. We introduce a new loss strategy where losses for basecalling and classification are back-propagated separately, with model weights combined for the shared layers, and a pre-configured ranking strategy allowing top-K species accuracy, giving users flexibility to choose between higher accuracy or higher speed at identifying the species. We achieve state-of-the-art basecalling accuracies, while classification accuracies meet and exceed the results of state-of-the-art binary classifiers, attaining an average of 92.5\%/98.9\% accuracy at identifying the top-1/3 species among a total of 17 genomes in the Wick bacterial dataset. The work presented here has implications for future studies in metagenomic profiling by accelerating the bottleneck step of matching the DNA sequence to the correct genome. 
\end{abstract}

\section{Introduction}
As the cost of genome sequencing has fallen drastically over the last decade~\citep{accesswire2022}, genome sequencing has seen a rapid uptick in a range of fields such as forensics~\citep{maria2017}, crop analysis~\citep{cruz2023}, and metagenomic analysis~\citep{LaPierre2020}. In particular, metagenomics is crucial to understanding the wider context of a single genome within its community~\citep{Handelsman2004}. Metagenomic profiling is the process by which relative abundance of a genome within a sample is ascertained~\citep{Quince2017}.

Metagenomic profiling can be accomplished by a variety of alignment-based~\citep{LaPierre2020, Bağcı2021} and non-alignment-based methods~\citep{Wood2019}~\citep{Ounit2015}, which vary in their computational complexity and memory consumption~\citep{LaPierre2020}, as well as accuracy by various metrics~\citep{McIntyre2017}. Critically, non-alignment-based methods have been demonstrated to have a significant trade-off between precision (false-positive rate) and recall (false-negative rate)~\citep{Sczyrba2017}. In contrast, alignment-based methods have been demonstrated to perform better in both metrics at the cost of drastically increased memory and computational requirements~\citep{LaPierre2020}, as each read must be aligned against a vast database of possible genomes. Methods for reducing these runtime requirements while maintaining high profiling accuracy are thus attractive research opportunities.

In parallel to the above research, pre-basecalling analysis is being pursued by many research groups. Basecalling is the process which precedes alignment, during which a raw signal sample, hereafter referred to as a read, emanating from a sequencing device, for example Oxford Nanopore Technology's (ONT) MinION sequencer~\citep{wang2021}, is translated into a chain of nucleotide bases. Inferring these k-mer sequences can be accomplished via Deep Neural Networks (DNNs), which provide State of the Art (SotA) speed and accuracy in comparison to previous methods~\citep{Wick2019}. Despite this, the basecalling step is still commonly the bottleneck in the analysis pipeline, consuming up to 40\% of runtime even running on a GPU~\citep{lou2020}. Therefore, much research has focused on eliminating unnecessary basecalling on non-target reads by identifying and rejecting them early~\citep{cavlak2022, dunn2021, kovaka2021}. Most of these methods rely on DNNs to identify a single target genome, i.e. human, and reject all others, making them binary classifiers.

An intuitive extension of these read classifiers would be to move towards multi-class classification, where a read is categorized amongst a pool of possible genomes, rather than a simple binary classification. This extension allows for either precise classification of each read or the generation of a candidate list of genomes for comparison in the post-basecalling classification step. By narrowing down the possible genomes early in the process, computational requirements are significantly reduced compared to traditional metagenomic profiling methods.

In this work, we present the first multi-objective deep neural network for multi-class classification and basecalling to reduce downstream alignment overhead. To accomplish this:
\begin{figure*}
\centerline{\includegraphics[width=0.9\textwidth]{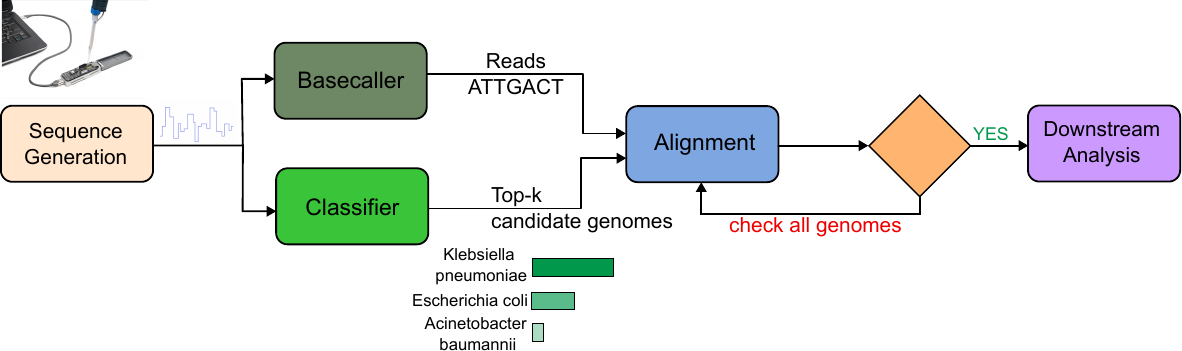}}
\caption{Proposed genome sequencing pipeline with species classification while basecalling.}
\vspace{-15pt}
\label{figA}
\end{figure*}
\begin{itemize}
    \item We augment the traditional Bonito DNN basecaller~\citep{wright2020} with a classification layer, enabling classification while basecalling. We explore two model variants, namely parallel and serial approaches with respect to the basecaller decoder. 
    \item %
    \rone{We develop a custom loss strategy} that combines basecalling and classification losses, giving more weight to classification predictions made later in the process when the model has seen more sequence data and is more confident.
    \item We propose a pre-configured ranking strategy, where the top-K predicted classes are passed to the next stages of the genome sequencing pipeline, allowing flexible trade-offs between accuracy and computational efficiency. A consensus-based testing metric is used to assess the final classification accuracy.
    \item We train our network on a set of 17 genomes, demonstrating between 92.5\%/98.89\% top-1/top-3 per-read classification accuracy without degrading basecalling accuracy.
\end{itemize}

The rest of this paper is organized as follows. Section~\ref{sec:background} provides background, while Section~\ref{sec:preclassification} gives details on our pre-classification strategy. Then Section~\ref{sec:setup} and Section~\ref{sec:results} detail our experimental setup and results. Section~\ref{sec:integration} explains the integration of our solution in metagenomic profiling pipelines. Finally, Section~\ref{sec:conclusion} concludes this work.

\section{Background}
\label{sec:background}
For extracting nucleotide sequences of and performing analysis on DNA/RNA samples, Oxford Nanopore Technologies (ONT)~\citep{ONT} offers state of the art devices based on the usage of flow cells which consist of nanoscopic pores integrated in an electrically-resistant polymer membrane. Molecules of the different samples pass through these nanopores and produce an electrical change in the current that is detected by the electrode and sensor corresponding to the nanopore. The MinION device is able to produce 0.46 GB of raw electrical signal per minute with samples millions of bases long~\citep{minion}, greatly enhancing downstream accuracy~\citep{ONT} after passing through the downstream processing steps.
\subsection{Basecalling Algorithms}
Basecalling algorithms are responsible for converting the raw electrical signal into  sequences of nucleotide bases representing the original DNA or RNA molecule. Basecalling originally utilized heuristics, statistical methods, or direct measurements~\citep{Sanger1977-in}~\citep{CANARD19941}~\citep{Rusk2011}, but they were limited in terms of size of datasets, noise handling, and pattern complexity. With the rise of DNNs, scalability, and more efficient handling of large datasets, recognition of complex patterns became more feasible and DNN-based basecallers were developed, offering superior performance to prior approaches~\citep{Heather2015-vr}. 

The typical ONT workflow for the basecalling task consists of the basecalling DNN and a decoder. The DNN commonly consists of networks such as CNNs acting as feature extractors and an inference trunk, such as LSTMs or transformers, with a final fully connected layer. The established decoder in literature is a hybrid Conditional Random Field (CRF)/Connectionist Temporal Classifier (CTC), which handles well the time variant nature of the raw electrical signals~\citep{CTC-CRF}. Some of the most well-known DNN-based basecallers include Guppy~\citep{guppy_protocol},
Bonito~\citep{wright2020}, Mincall~\citep{Miculinic2019-hh}, Causalcall~\citep{Zeng2019-bc}, Halycon~\citep{10.1093/bioinformatics/btaa953}, CATCaller~\citep{Lv2020-uy}, URNano~\citep{Zhang2020-op} and SACall~\citep{Huang2022-gx}. 
In our study, we utilize the Bonito model within the framework described in~\citep{paga_basecalling_architectures}, since this model is the standard provided by ONT and demonstrates high basecalling performance based on the metrics outlined in the framework's associated paper~\citep{Pages-Gallego2023-jg}.
\subsection{Genome Classification}
Classification is the process of identifying to which genome a read belongs. Typically, this is performed post-basecalling; however, several studies have explored the possibility of pre-basecalling classification. Specifically, binary classification is studied across a variety of works, often exploring algorithms to support the ‘‘Read-Until'' (i.e. early read ejection) option of the nanopore sequencer. This feature refers to the capability of the sequencing device to discard a partially sequenced molecule deemed off-target~\citep{nanoporetechReadUntil}. 
SquiggleNet~\citep{Bao2021-kf} is the first deep-learning method which can directly classify DNA sequences from electrical signals. It features a 1D-ResNet-inspired architecture, using bottleneck convolutional layers 
followed by 
a final fully connected layer for classification. SquiggleNet achieves over 90\% accuracy in distinguishing human from bacterial DNA, and generalizes to new bacterial species in respiratory samples.
DeepSelectNet~\citep{Senanayake2023-sl} is the refined successor of SquiggleNet, achieving an approximately 12\% improvement in accuracy via enhanced training data preprocessing and improved feature extraction, reaching an accuracy of over 90\% across 5 different datasets while addressing some of the limitations of SquiggleNet.
TargetCall~\citep{cavlak2022} is a binary classification tool 
that 
serves as a pre-basecalling filter. It processes complete sequences 
from the nanopore device and classifies them as on- or off-target based on the reference genome of interest. 
It utilizes two steps, the first being a lightweight DNN based on Bonito, which produces low accuracy sequences that are nevertheless sufficient for detecting if they belong to the target reference genome. \rthree{This is followed by a block which performs similarity check using minimap2}~\citep{minimap2}.
TargetCall reports 98.88\% sensitivity in keeping on-target reads and up to 94.71\% filtering out of off-target reads. 

\rone{While the aforementioned works perform well in applications where binary classification is sufficient, tasks such as metagenomic profiling where the identification of the genome of interest requires comparisons to more than one species would benefit from a multi-class classifier, as this work presents.}
\subsection{Metagenomic profiling}
The objective of the domain of metagenomic profiling 
is to study and understand microbial communities by estimating the relative abundance of taxa in a sample of various species. Traditional methods, which used culture-based analysis, were later replaced by techniques involving high-throughput sequencing~\citep{LaPierre2020}. The reads generated from these techniques go through the step of classification, which bins them into organism groups. 
Different approaches appear in literature, but there is an inevitable trade-off between false positive rate (precision) and false negative rate (recall). This is true for both alignment- and non-alignment-based methods~\citep{LaPierre2020}. Among the alignment-based methods, Metalign~\citep{LaPierre2020} has achieved a good balance between precision and recall, while accelerating the classification step by reducing the database of genomes to align against by up to 100x. 
The most famous non-alignment-based approach is Kraken~\citep{Wood2014-st}, which offers low latency classification at the cost of lower accuracy compared to alignment based methods~\citep{LaPierre2020}. 

In summary, while alignment-based profiling offers a better balance between precision and recall, it is computationally intensive. In contrast, non-alignment methods, though faster, often sacrifice accuracy. Our work aims to enhance the efficiency of alignment-based approaches by accelerating them while maintaining their high accuracy, but its usage can be extended to non-alignment methods to improve their performance as well.

\section{Pre-Classification while basecalling}
\label{sec:preclassification}
\begin{figure}
\centerline{\includegraphics[width=0.9\textwidth]{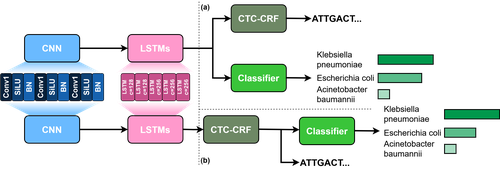}}
\caption{Proposed parallel (a) and serial (b) models for the task of classification while basecalling.}
\label{figB}
\vspace{-15pt}
\end{figure}
The novelty of this work is to predict during the basecalling step, i.e. during sequencing, to which species a read belongs, as illustrated in Fig.~\ref{figA}. By having a preliminary candidate or candidates for classification, the database to which the read must be matched can be reduced to 1 or top-K species, reducing classification latency and computational overhead. 
Metalign previously demonstrated how pre-filtering the database against which reads are aligned improves throughput while maintaining balance between precision and recall~\citep{LaPierre2020}, a concept this paper extends. By reducing the processing time of this method, the entire pipeline of metagenomic profiling will be enhanced, raising the standard for this and other methods in terms of the accuracy vs. latency trade-off. 
\subsection{DNN-based basecaller model architecture}
The proposed model architecture is based on the existing Bonito~\citep{wright2020} basecaller, but can be applied to any DNN basecalling network. Bonito, illustrated in Fig.~\ref{figB}, consists of 3 convolutional layers followed by 5 LSTM layers and a fully connected layer $F_0$.
The stride of the last layer of the CNN is 5; this translates to e.g. 800 timesteps for the sequence when the window-size of the input signal is 4,000. The CRF decoder is fed the output of the fully connected layer to produce the sequence of nucleotides. 
The classification portion of the network consists of a newly added fully connected layer whose output size is equal to the number of species to be classified against.
\subsection{Model architectures for classification while basecalling}
 While the Bonito model is employed for executing the basecalling step, the framework from~\citep{paga_basecalling_architectures} \rone{is extended in two ways by choosing where to add the aforementioned classification fully connected layer, either in parallel or in series with} $F_0$. In the parallel approach, Fig.~\ref{figB}(a), the backpropagation in the classifier is independent from the basecaller's decoder. In the second approach displayed in Fig.~\ref{figB}(b), the loss of the classifier includes the decoder, creating a dependent relationship between them during the backward pass. The input size of the classification layer is equal to either the output size of the LSTMs (384) if it is implemented in parallel with $F_0$, or the output size of $F_0$ if implemented in series with $F_0$. 

\subsection{Training for basecalling/classification}
\rone{During the training and optimization of the network, separate losses are calculated for the CTC-CRF block and the classifier block. Basecalling loss is calculated from the CRF/CTC decoding via the} \textit{seqdist} library~\citep{seqdist}.
For the classification loss, CrossEntropy is applied without reduction, maintaining individual loss values for each element. The obtained result contains loss values for all the time-steps, representing the prediction of the species at each base. Since the model is necessarily less confident of its prediction at earlier time-steps in comparison to later, a scaling factor between 0 and 1 is applied to each time-step, with the average of the scaled loss across all time-steps taken as the final loss. Explorations were made with various scaling factor functions, with the logarithmic scaling factor resulting in best performance.

Basecalling and classification loss are back propagated independently through their respective linear layers, then the gradients they produce are summed and back propagated through layers shared by both loss contributors. The classifier and the decoder thus contribute equally to the CNN and LSTM portions of the network in both aforementioned parallel and serial implementations. In contrast, in the parallel implementation, the fully connected layers are updated independently of each other, while for serial implementation, the classifier loss is back propagated through $F_0$ as well. Section~\ref{subsec:modelcompare} discusses the accuracy impact of the parallel vs. series architectures.

For evaluating the model during the training process, a validation check is done every 500 batches. For calculating the basecalling accuracy, an alignment score is calculated for each basecalled read using the parasail library~\citep{parasail}. For calculating classification accuracy, parametric top-K MulticlassSo Accuracy~\citep{multiclass} from PyTorch is used. The prediction in the last time step is passed to that metric, together with the correct label of the species. Different values of \textit{K} are studied to analyze the trade-off between prediction accuracy and reduced computational complexity during downstream analysis.
\subsection{Data processing for classification}
\label{subsec:balance}
As the task is to differentiate between different species, we found that a balanced dataset between the target species greatly improved accuracy. As lengths of each read in an ONT dataset may vary significantly, it is necessary to number the samples in each read and balance the dataset according to sample count. Each is then split into chunks which are passed through the basecaller. We also shuffled the genomes in the training set so that the network trains on classes in a homogeneous manner. Preprocessing of the data to prepare it for training is described in greater detail in Section~\ref{subsec:dataprep}.
\section{Training/test set and experimental setup}
\label{sec:setup}
We utilize the popular Wick dataset for experimental analysis of our basecalling/classification network~\citep{wick2019_2}. We utilize the data preparation strategy presented in~\citep{paga_nanopore_benchmark} to build the training and validation sets. As species classification requires a balanced dataset as described in Section~\ref{subsec:balance}, and the Wick dataset consists of many species of varying read counts, we first select the datasets with more than 5,000 reads. This results in a total of 30 datasets as listed in the Table~\ref{tab1}, where 17 of them are unique species. For each of them, 500 reads are set aside for testing, and the rest are available for training and validation. To maintain consistency during training in species classification, we randomly select one collection of \textit{Klebsiella pneumoniae}, namely \textit{Klebsiella pneumoniae-INF042}, from its 14 options listed in Table~\ref{tab1}. The other 13 \textit{Klebsiella pneumoniae} sets are discarded from training, but their testing reads are used. 

We use the Wick dataset due to its popularity among researchers and its open-source nature. As this dataset utilizes ONT's R9 chemistry, its basecalling accuracy is not comparable to that of R10 chemistry which boosts accuracy to over 99\%~\citep{ont_news} but does not currently have widely available comprehensive open source datasets. Importantly, our methodology is not tied to the Wick dataset and can be extended to datasets using the latest flow cell chemistry without loss of generality.

Appendix~\ref{data_preprocessing} describes the preprocessing steps taken to prepare the data for training and inference, while appendix~\ref{experimental-setup} describes the hardware setup and training hyperparameters.

\section{Results}
\label{sec:results} %
\subsection{Parallel vs. Serial model architecture classification accuracy studies}
\label{subsec:modelcompare}
\begin{figure}
\centerline{\includegraphics[width=0.8\textwidth]{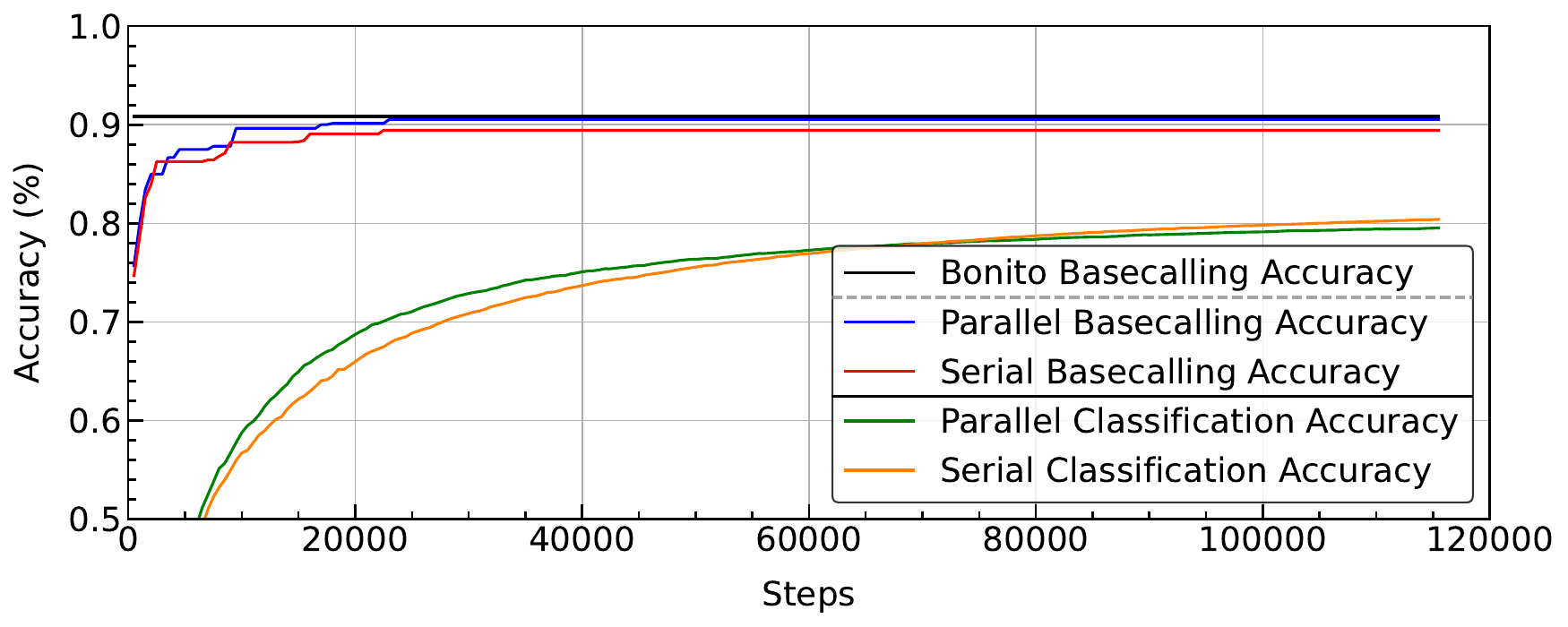}}
\vspace{-10pt}
\caption{Validation accuracies for basecalling and top-1 classification for the parallel and serial model architectures.}
\label{figC}
\end{figure}
Fig.~\ref{figC} shows the top-1 classification accuracies of the parallel and serial model architectures over 17 epochs. Classification accuracies using both architectures are similar, around $\sim$80\%. \rthree{While the parallel model architecture improves faster in the first 40,000 steps, both networks converge towards identical values.} We thus analyze the impact on basecalling accuracy to differentiate the networks.
\subsection{Impact on basecalling accuracy}
\begin{figure}
\centerline{\includegraphics[width=0.7\textwidth,trim={1cm 1cm 1cm 1cm},clip]{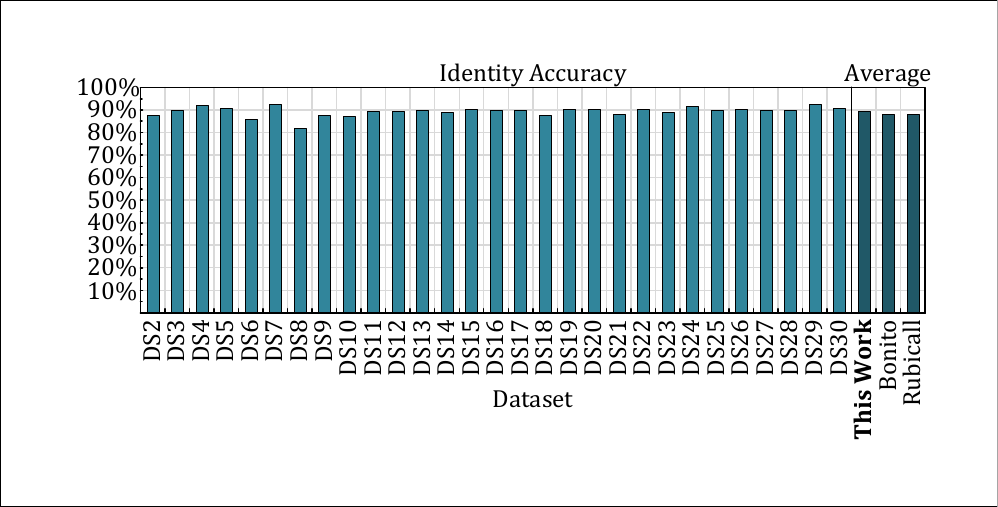}}
\caption{Post-alignment identity accuracy of this work vs. Bonito and SotA classifier RUBICALL~\citep{Singh2024-es}. }
\label{figD}
\vspace{-10pt}
\end{figure}
We compare the basecalling validation accuracies of the new models during 15 epochs with the original Bonito basecaller model trained on the same dataset in Fig.~\ref{figC}(b). Both model architectures exhibit similar trends, however it is observed that parallel network approaches the baseline accuracy within \textless0.5\%. The serial basecalling accuracy suffers as its fully connected layer is affected by the loss of the classifier output, while classifier accuracy does not significantly benefit from the extra layer in its pipeline. 
\rthree{Therefore, given that both configurations demonstrate similar classification accuracies and negligible runtime overhead as discussed in}~\ref{experimental-setup}, \rthree{the rest of this work utilizes the parallel model, noting that either are viable implementations for the task under consideration.} %

For downstream accuracy analysis setup, the framework proposed in~\citep{Singh2024-es} is utilized, namely, minimap2~\citep{minimap2} is used to align each read to its source genome. Fig.~\ref{figD} reports the identity accuracies for each dataset, the indices of which correspond to Table~\ref{tab1}, with dataset one left off due to failure to align. The average classification accuracy is comparable to both the standard Bonito model and the RUBICALL model, an SotA model demonstrated to outperform many previous advanced models on the same Wick dataset~\citep{Singh2024-es}. Additionally, when comparing common datasets in both works, it was observed that the results were generally consistent, with only minor differences of 1–2\%. These results indicate that the addition of the classifier layer does not impact basecalling accuracy.

\subsection{top-K per-chunk accuracy}
\label{subsec:per-chunck} 
\begin{figure}
\centerline{\includegraphics[width=0.8\textwidth]{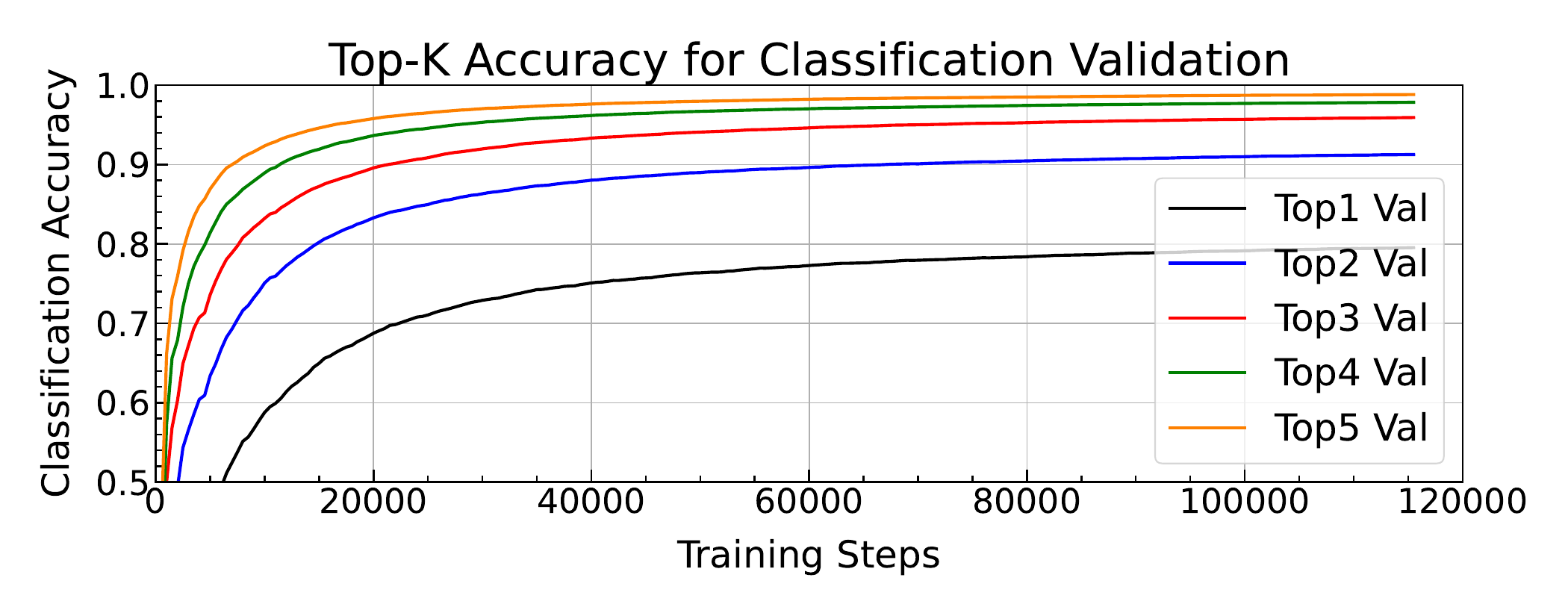}}
\vspace{-15pt}
\caption{\rfour{Top-K classification accuracy evolution during training of parallel model architecture.}}
\vspace{-5pt}
\label{figE}
\end{figure}
Fig.\ref{figE} illustrates the evolution of top-K classification accuracy during training for the parallel model architecture. It can be observed that top-1 classification saturates around 80\%, while it approaches 99\% as the top-K is increased up to 5. This configuration of top-K accuracy indicates flexibility when integrating the classifier model in downstream pipelines, as discussed in Section~\ref{sec:integration}. \rfour{We note that the choice of k does not necessitate retraining of the network and can be chosen after training dependent on downstream pipeline requirements.}
\subsection{Per-read classification accuracy}
\begin{figure}
\centerline{\includegraphics[width=1\textwidth]{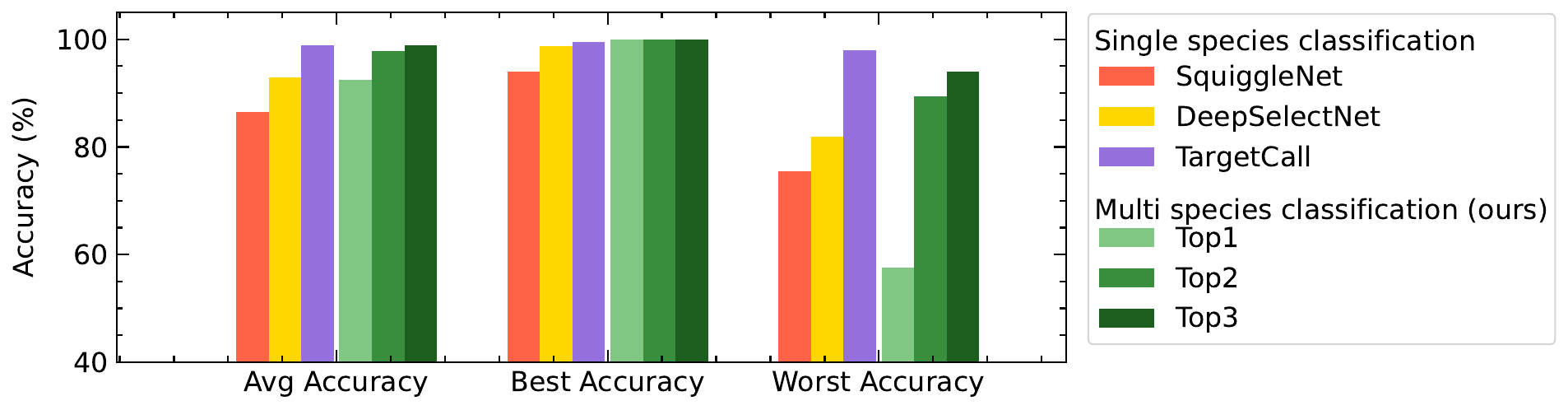}}
\vspace{-5pt}
\caption{\rone{The proposed basecaller/classifier model acheives SotA classification accuracy even while classifying between multiple species, against the single species classification of the SotA works.}}
\label{figF}
\vspace{-10pt}
\end{figure}
While Section~\ref{subsec:per-chunck} reports per-chunk classification accuracy, classification of an entire read consisting of multiple chunks is determined by a consensus-based approach, where predictions are made for individual chunks of each read, and the read is classified as the species with the highest vote amongst its chunks. The top-1 overall accuracy is calculated as the ratio of correctly classified reads to the total number of reads, using 500 reads per species for evaluation. For calculating top-K accuracy, top-K species with highest number of classified reads are included in the ratio, if the correct species is amongst those top-K species. The results for K=1-3 of this approach for the 17 unique species included in both training and testing are displayed in Fig.~\ref{figF}, alongside SotA binary classifiers. \rone{On average our model's multi-class accuracy meets and exceeds that of single-class networks SquiggleNet and DeepSelectNet in all configurations and matches TargetCall with top-3 classification.} %

\begin{figure}
\centering
\includegraphics[ width=0.7\textwidth]{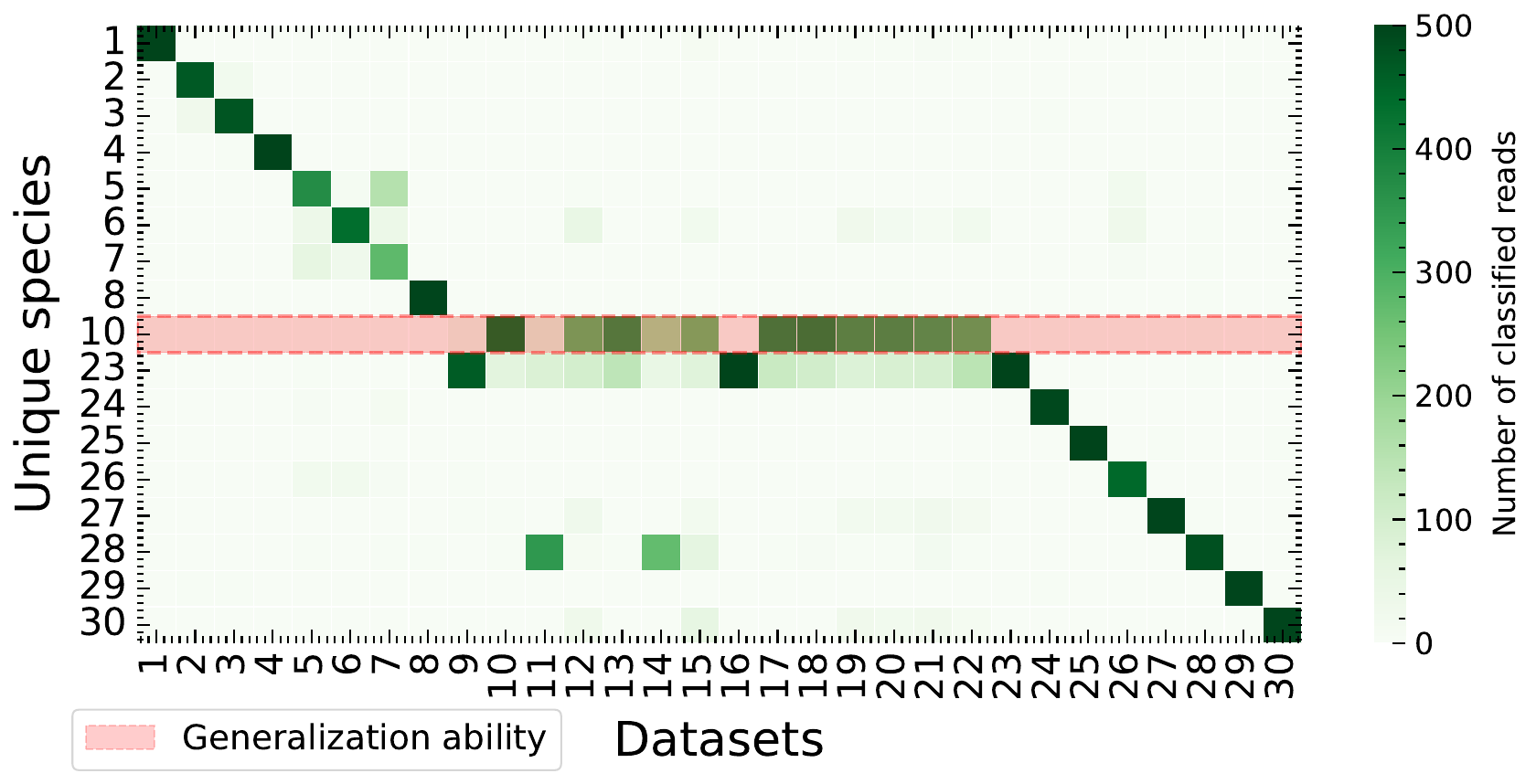}
\vspace{-10pt}
\caption{Read classification heatmap for datasets in Table~\ref{tab1}. Datasets of same species classify with high accuracy to training class of the same species.}
\label{figG}
\vspace{-10pt}
\end{figure}

The heatmap in Fig.~\ref{figG} illustrates how the 30 datasets in Table~\ref{tab1} are classified amongst the 17 unique training species. Clear diagonals on the left and right of the figure are noticeable, showcasing the accuracy of our model in differentiating unique species. In the center of the figure, it can be seen that the majority of the datasets belonging to \textit{Klebsiella~pneumoniae} converge towards the class of \textit{Klebsiella~pneumoniae}. This demonstrates our model's generalizability for classifying datasets unseen in the training set. Poorly classified species, namely, \textit{Escherichiaq marmotae}, incorrectly classified 31\% of the time as \textit{Citrobacter freundi}, and \textit{Klebsiella pneumoniae} (datasets 11 and 14), which are misclassified as \textit{Shigella sonnei}, can be attributed to the similarity between these genomes, as for example the Jensen-Shannon divergence between the 9-mer relative counts of \textit{Escherichia marmotae} and \textit{Citrobacter freundi} is less than 0.1~\citep{Pages-Gallego2023-jg}. This suggests further research into gene family classification for highly similar genes.

Nevertheless, our model achieves SotA classification accuracy while also extending the functionality from binary to multi-class classification, all within the original basecalling framework and without introducing a classifier-specific DNN.

\section{Integration in metagenomic profiling pipelines}
\label{sec:integration}
While this work focuses primarily on the accuracy of our proposed method, we provide here some insight into how it may be integrated into the wider metagenomic classification pipeline. This pipeline faces a challenge in that, while basecalling and assembly computational overhead do not scale with number of species, classification computational requirements scale as the genome database grows. Metalign demonstrates a reduction of up to 100x on number of genomes against which to match a given set of samples for a database of 199,807 microbial genome assemblies compiled from the RefSeq~\citep{refseq} and GenBank~\citep{genbank} database, with a comparable reduction in alignment time. Even so, a reduction of 100x still results in $\sim$2000 genomes to which each read must be aligned. \rthree{While we initially study here a relatively small database of 17 species to understand the feasibility of multi-class read classification, we plan to expand the study by developing a training dataset containing larger numbers of genomes, and classifying to families of genomes.} 

The method proposed here reduces the number of alignments that must be made for each read to the top-K most likely candidates while maintaining alignment accuracy as, if the network misclassifies a read resulting in no or poor alignment, the read can be re-aligned against the comprehensive genome database. This motivates an interesting research avenue of exploring optimal top-K values to balance the trade-off between the number of network misclassifications against the necessity of aligning against more candidate genomes. This strategy most benefits alignment-based classifiers, who suffer more from the expensive computational alignment step, but also applies to alignment-free classifiers. 
\section{Conclusion}
\label{sec:conclusion}
In summary, this study demonstrates how a DNN-based basecaller like Bonito can be expanded with a classification layer in two possible architectures, parallel and serial. A tailored loss method is developed which encapsulates the basecalling and classification loss. While for basecalling, the prediction of bases at each time-step contributes equally to the loss calculation, for classification, the predictions in the later stages carry more weight than the initial ones. The model is trained on a set of 17 genomes. During testing, for the generated sequences, an alignment score is produced using a read mapper and their reference genome, and the classification accuracy is obtained using a consensus-based approach which is implemented to produce a per-read prediction from the per-chunk predictions. Both of the tasks prove to be successful in achieving high accuracies, e.g. 90\% for the basecaller and an average accuracy of  92.5\% for top-1 classification and 98.89\% for top-3 classification. These classification results will help speed up species identification in the metagenomic profiling pipeline by reducing the amount of required  genome comparisons.

\section*{Acknowledgements}
This work was supported by European Union’s Horizon Europe Research and Innovation Program (BioPIM, Grant 101047160), and Swiss State Secretariat for Education, Research and Innovation (SERI) (Grant 22.00076).

\bibliographystyle{iclr2025_conference}
\bibliography{bibliography}

\appendix
\section{Appendix}
\subsection{Data Preprocessing}
\label{data_preprocessing}
As the Wick dataset does not provide ground truth nucleotide sequences for most of its data, it is necessary to generate these sequences for the training set. For each species there are \textit{fast5} files~\citep{Gamaarachchi2022} containing raw electrical signals (reads), and reference genomes given as \textit{fna} or \textit{fasta} files, and for some of them there are \textit{fastq} files which represent the ground truth of the nucleotide sequence. \textit{fastq} files containing inferred nucleotide sequences are generated for each species using the \textit{dorado dna\_r9.4.1\_e8\_sup@v3.6} basecalling network~\citep{ONT}. The original reads are then annotated with the generated files and ``resquiggled'' using their corresponding reference files (\textit{fna} or \textit{fasta}). The resquiggle process refers to the correction of basecalling errors by re-assigning the nanopore reads to a reference sequence~\citep{tombo}. After these steps, the reads of each species are divided into a ratio of 3:1 training/validation sets, with each read divided into ``chunks'' of signals of window-size 4,000. As reads contain a widely varying number of signal values, up to 3x difference in amount of total chunks per species, it is necessary to balance the dataset at a chunk granularity. Thus, the number of chunks included for each species is limited to that of the species with the least number of chunks. This species is \textit{Pseudomonas\_aeruginosa-MINF\_7A}, consisting of 68k chunks, or $\sim$2.03 GB.
The complete training and validation datasets are then shuffled so the network learns to classify all species in parallel.
Each chunk in the final dataset consists of the original sample, the ground truth according to the resquiggling process, and a classification index between 0 and 16, corresponding to a unique species shown in Table~\ref{tab1}. %

\begin{table}
\caption{Datasets and their Read Counts for the experiments}
\footnotesize
\begin{center}
\begin {tabular}{r|l|c}
\hline
\textbf{Index} & \textbf{Name of species} & \textbf{Nr. of Reads} \\
\hline\hline
1 & Acinetobacter\_baumannii-AYP\_A2 & 6558\\
\hline
2 & Acinetobacter\_nosocomialis-MINF\_5C & 6722\\
\hline
3 & Acinetobacter\_ursingii\_MINF\_9C  & 6976\\
\hline
4 & Burkholderia\_cenocepacia-MINF\_4A & 7096\\
\hline
5 & Citrobacter\_freundii-MSB1\_1H & 7093\\
\hline
6 & Escherichia\_coli-MSB2\_1A & 6985\\
\hline
7 & Escherichia\_marmotae-MSB1\_5C & 7064\\
\hline
8 & Haemophilus\_haemolyticus-M1C132\_1 & 8669\\
\hline
9 & Klebsiella\_pneumoniae-INF032 & 14320\\
\hline
10 & Klebsiella\_pneumoniae-INF042 & 10695\\
\hline
11 & Klebsiella\_pneumoniae-INF116 & 6776\\
\hline
12 & Klebsiella\_pneumoniae-INF215 & 7142\\
\hline
13 & Klebsiella\_pneumoniae-INF322 & 7212\\
\hline
14 & Klebsiella\_pneumoniae-KSB1\_1I & 7031\\
\hline
15 & Klebsiella\_pneumoniae-KSB1\_6G & 7040\\
\hline
16 & Klebsiella\_pneumoniae-KSB1\_7E & 5832\\
\hline
17 & Klebsiella\_pneumoniae-KSB1\_9A & 6787\\
\hline
18 & Klebsiella\_pneumoniae-KSB2\_1B & 16847\\
\hline
19 & Klebsiella\_pneumoniae-NUH11 & 7336\\
\hline
20 & Klebsiella\_pneumoniae-NUH27 & 7321\\
\hline
21 & Klebsiella\_pneumoniae-NUH29 & 15178\\
\hline
22 & Klebsiella\_pneumoniae-SGH07 & 5645\\
\hline
23 & Klebsiella\_variicola-INF022 & 6501\\
\hline
24 & Morganella\_morganii-MSB1\_1E & 6307\\
\hline
25 & Pseudomonas\_aeruginosa-MINF\_7A & 7082\\
\hline
26 & Salmonella\_enterica-21\_06152 & 6638\\
\hline
27 & Serratia\_marcescens-17\_147\_1671 & 11742\\
\hline
28 & Shigella\_sonnei-212\_0237 & 23583\\
\hline
29 & Staphylococcus\_aureus-CAS38\_02 & 11047\\
\hline
30 & Stenotrophomonas\_maltophilia-17\_G\_0092~~~~~ & 16010\\
\hline
\end{tabular}
\label{tab1}
\end{center}
\end{table}
\label{subsec:dataprep}

\subsection{Training setup} %
\label{experimental-setup}
The training is performed with an x86 architecture, 16-core CPU and a single \rone{NVIDIA Tesla V100 GPU} supported by 64GB of RAM. The implementation is performed using PyTorch 2.3, with CUDA 12.1 for GPU acceleration. Python 3.7 is employed, along with other dependencies mentioned in~\citep{paga_basecalling_architectures}, from which the training framework is retrieved. \rfour{The model is trained with a window size of 4,000 (Bonito default setting), window overlap of 0, and a batch size of 64 as dictated by GPU VRAM capacity. The initial learning rate is set to 0.01 with a warm-up phase of 1,000 steps}~\citep{Pages-Gallego2023-jg} \rfour{and is reduced using the \textit{ReduceLROnPlateau} LR strategy as loss converges.} \rthree{Training is conducted for 17 epochs until convergence, taking around 13 hours. We note that the addition of the classifier layer has negligible (\textless3\%) impact on training time in either series or parallel configuration.}

\end{document}